\documentclass[aps,preprint,preprintnumbers,amsmath,amssymb]{revtex4}
\usepackage{mathrsfs}
\usepackage{amssymb}
\usepackage{graphicx}% Include figure files
\usepackage{dcolumn}% Align table columns on decimal point
\usepackage{bm}% bold math
\usepackage{color}
\usepackage[breaklinks=true,colorlinks=true,linkcolor=blue,urlcolor=blue,citecolor=blue]{hyperref}
\usepackage{soul}
\usepackage{ulem}
\makeatletter
\newcommand{\rmnum}[1]{\romannumeral #1}
\newcommand{\Rmnum}[1]{\expandafter\@slowromancap\romannumeral #1@}

\begin{document}

\title{Unconventional and conventional quantum criticalities in CeRh$_{0.58}$Ir$_{0.42}$In$_5$}

\author{Yongkang Luo$^{1}$\footnote[1]{Electronic address: mpzslyk@gmail.com}, Xin Lu$^{2}$, Aadm P. Dioguardi$^{1}$, Priscila F. S. Rosa$^{1}$, Eric D. Bauer$^{1}$, Qimiao Si$^{3}$, and Joe D. Thompson$^{1}$\footnote[2]{Electronic address: jdt@lanl.gov}}

\address{$^1$Los Alamos National Laboratory, Los Alamos, New Mexico 87545, USA;}
\address{$^2$Center for Correlated Matter and Department of Physics, Zhejiang University, Hangzhou 310058, China;}
\address{$^3$Department of Physics and Astronomy and Center for Quantum Materials, Rice University, Houston, Texas 77005, USA.}

\date{\today}

%\pacs{75.20.Hr, 74.70.Tx, 74.62.Fj, 74.25.fg}
%75.20.Hr Local moment in compounds and alloys; Kondo effect, valence fluctuations, heavy fermions
%74.70.Tx Heavy-fermion superconductors
%74.62.Fj Effects of pressure;
%74.25.fg Thermoelectric effects
%71.27.+a Strongly correlated electron systems; heavy fermions

\maketitle

\textbf{
An appropriate description of the state of matter that appears as a second order  phase transition is tuned toward zero temperature, {\it viz.} quantum-critical point (QCP), poses fundamental and still not fully answered questions. Experiments are needed both to test basic conclusions and to guide further refinement of theoretical models. Here, charge and entropy transport properties as well as AC specific heat of the heavy-fermion compound CeRh$_{0.58}$Ir$_{0.42}$In$_5$, measured as a function of pressure, reveal two qualitatively different QCPs in a {\it single} material driven by a {\it single} non-symmetry-breaking tuning parameter. A discontinuous sign-change jump in thermopower suggests an unconventional QCP at $p_{c1}$ accompanied by an abrupt Fermi-surface reconstruction that is followed by a conventional spin-density-wave critical point at $p_{c2}$ across which the Fermi surface evolves smoothly to a heavy Fermi-liquid state. These experiments are consistent with some theoretical predictions, including the sequence of critical points and the temperature dependence of the thermopower in their vicinity.
}\\

\textbf{INTRODUCTION}

Heavy-fermion metals have emerged as prototypes for discovering quantum-critical states\cite{Coleman-QCP2005,Gegenwart2008} that are of broad interest as they are believed to be the origin of non-Fermi-liquid (NFL) and unconventional superconducting (SC) phases in classes of strongly correlated electron materials, ranging from organics to metallic oxides. Generically, a QCP is an end point at absolute zero temperature of a continuous transition that separates ordered and disordered phases and is accessed by a non-thermal control parameter $g$, such as chemical doping ($x$), pressure ($p$) and magnetic field ($B$)\cite{Sachdev-QPT,Lohneysen-RMP2007}. The conventional model of quantum criticality is based on a quantum extension of the Landau-Ginzburg-Wilson theory of classical, thermally-driven phase transitions and considers only fluctuations of a spin-density-wave (SDW) order parameter\cite{Lohneysen-RMP2007}. In this model, which does not treat electronic degrees of freedom as part of the critical excitations, the Fermi surface evolves smoothly as a function of $g$ across the QCP\cite{Hertz-QCP,Millis-QCP}. Though this model provides a reasonable account of physical properties in some systems near a QCP\cite{Lohneysen-RMP2007}, it fails fundamentally to describe critical responses in other metallic systems in which there is accumulating evidence for unconventional quantum criticality, most notably in heavy-fermion compounds\cite{Schroder-CeCu6AuQCP,Custers-YbRh2Si2QCP,Friedemann-YbRh2Si2_CoIr,Custers-Ce3Pd20Si6QCP,Paschen-YbRh2Si2Hall,LuoY-CeNiAsOQCP, LuoY-CeNi2As2Pre}. Alternatives to the conventional model, frequently called local, selective Mott or Kondo-breakdown theories, invoke criticality of electronic degrees of freedom that may be concurrent with magnetic criticality\cite{SiQ-localQCP,Coleman-QCP2005,Gegenwart2008,Pepin-Mott2008}, and the QCP is accompanied by a sharp reconstruction of the Fermi surface (FS). These models of criticality are unconventional in that they go beyond the Landau-Ginzburg-Wilson framework. Though there is experimental support for them\cite{Schroder-CeCu6AuQCP,Custers-YbRh2Si2QCP,Friedemann-YbRh2Si2_CoIr,Custers-Ce3Pd20Si6QCP,Paschen-YbRh2Si2Hall,LuoY-CeNiAsOQCP,LuoY-CeNi2As2Pre}, the concept of fluctuations of a symmetry-breaking order parameter with qualitative reconstruction of electronic states requires further theoretical understanding. For progress, it is important for experiments to both test their basic conclusions, such as the evolution of the Fermi surface across the QCP, and to guide their further development.

The critical state that develops near the $T$=0 magnetic/non-magnetic boundary as a function of $x$ in heavy-fermion CeCu$_{6-x}$Au$_x$ motivated early models of unconventional QCPs under applied pressure or magnetic field\cite{Schroder-CeCu6AuQCP}. Similar to CeCu$_{6-x}$Au$_x$ under applied pressure and magnetic field\cite{Lohneysen-CeCu6Au2001}, two very different critical states are realized in CeRhIn$_5$ when different tuning parameters are used to access its QCPs. Applying pressure to CeRhIn$_5$ induces in the limit $T$$\rightarrow$0 a magnetic/non-magnetic transition at a critical pressure $P_2$=2.4 GPa\cite{Knebel-CeRhIn5Pre,Park-CeRhIn5QCP,Park-CeRhIn5NJP}, where deHaas-vanAlphen (dHvA) measurements find an abrupt change from small-to-large Fermi surfaces and strong enhancement of the quasiparticle effective mass\cite{Shishido-CeRhIn5dHvA}, characteristic of an unconventional QCP. In the absence of an applied magnetic field, these responses are hidden by a dome of superconductivity (SC) that also encompasses a range of lower pressures where antiferromagnetic (AFM) order and SC coexist. At atmospheric pressure but as a function of magnetic field, which unlike pressure breaks time-reversal symmetry, there is a small-to-large Fermi surface reconstruction near 30 T within an AFM phase\cite{JiaoL-CeRhIn5B,Moll-CeRhIn5B} that terminates in a SDW-type QCP at $\sim$50 T\cite{JiaoL-CeRhIn5B}.
%there are two field-induced QCPs, one near 30 T where there is a change from small-to-large Fermi surface within an AFM phase\cite{JiaoL-CeRhIn5B,Moll-CeRhIn5B} and a SDW-type QCP near 50 T\cite{JiaoL-CeRhIn5B}.

The evolution of magnetism and SC in the homologous series CeRh$_{1-x}$Ir$_x$In$_5$ as a function of Ir content $x$ is somewhat similar to CeRhIn$_5$ under pressure.
%A quantitative scaling law estimates that CeIrIn$_5$ is under an effective chemical pressure of about 10 GPa with respect to CeRhIn$_5$\cite{Nicklas-CeRhIn5_Ir}.
For $x$$<$0.3, there is only large-moment, incommensurate AFM order at $\mathbf{Q}$=(0.5, 0.5, $\sim$0.297)\cite{Bao-CeRhIn5Neu}, and for 0.3$<$$x$$<$0.6, an additional small-moment, commensurate (0.5, 0.5, 0.5) AFM order develops and coexists with SC\cite{Llobet-CeRhIn5_IrNeu}. At higher Ir concentrations, there is only SC\cite{Pagliuso-CeRhIn5_Ir,Nicklas-CeRhIn5_Ir}. Interestingly, as in CeRhIn$_5$ under pressure\cite{Aso-CeRhIn5Neu}, the appearance of SC in CeRh$_{1-x}$Ir$_x$In$_5$ coincides with a change of magnetic structure\cite{Llobet-CeRhIn5_IrNeu}, making the similarity between these two cases even stronger. dHvA measurements show that the cyclotron frequencies are larger in CeIrIn$_5$ than in CeRhIn$_5$\cite{Shishido-Ce115FS}, which implies that cerium's $4f$ electron participates in making the larger FSs of CeIrIn$_5$ but is localized in CeRhIn$_5$. Somewhere in the CeRh$_{1-x}$Ir$_x$In$_5$ series, there should be a change in $4f$-electron localization and a magnetic QCP. Unfortunately, dHvA measurements have not been possible across the series. To explore these issues, we have measured the effect of pressure on the resistivity, thermopower and AC specific heat of a crystal of CeRh$_{0.58}$Ir$_{0.42}$In$_5$ whose nominal composition places it close to the critical Ir concentration $x_c$$\simeq$0.6\cite{Pagliuso-CeRhIn5_Ir} where the magnetic boundary drops rapidly towards $T$=0 at atmospheric pressure\cite{Llobet-CeRhIn5_IrNeu}. As will be discussed, these experiments reveal signatures that point to two distinct types of criticality as a function of a single, non-symmtery-breaking tuning parameter, pressure, and provide an important test of some theoretical predictions.\\

\textbf{RESULTS}

Figure~1(a) shows the temperature dependence of in-plane resistivity of CeRh$_{0.58}$Ir$_{0.42}$In$_5$ at pressures up to 2.20 GPa. At ambient pressure, $\rho(T)$ initially decreases upon cooling but starts to increase below 135 K and forms a pronounced broad peak around $T_{max}$=12 K, characteristic of the onset of coherent Kondo scattering. Below $T_{max}$, $\rho(T)$ decreases monotonically, and the sample becomes a superconductor with an onset critical temperature $T_{c}^{on}$=1.04 K. Above $T_{c}^{on}$, an inflection in $\rho(T)$ near $T_N$=3.5 K reflects a reduction of spin-scattering due to the formation of long range AFM order of Ce moments\cite{Pagliuso-CeRhIn5_Ir,ZhengG-CeRhIn5_IrNQR,Nicklas-CeRhIn5_Ir,Llobet-CeRhIn5_IrNeu}. Under pressure, (\rmnum{1}) $T_{max}$ increases monotonically and reaches $\sim$50 K at 2.20 GPa; (\rmnum{2}) $T_c^{on}$ increases with pressure and reaches a maximum of 1.45 K at $p$=1.06 GPa before it starts to decrease [Fig.~1(b)]; and, (\rmnum{3}) the AFM order is gradually suppressed and becomes unresolvable when $p$$>$0.48 GPa where the resistive anomaly is concealed by SC.

The zero-field pressure-temperature phase diagram of CeRh$_{0.58}$Ir$_{0.42}$In$_5$, constructed from resistivity measurements, is summarized in Fig.~1(c) where we see that evidence for magnetic order ($T_N$) disappears near 0.48 GPa below a dome of SC. A field of 9 T
completely suppresses SC but has a negligible effect on the N\'{e}el temperature and the temperature dependence of resistivity [inset to Fig.~1(a)], implying that no significant magnetic-structure or Fermi-surface change is induced by this modest field. In this field, however, a resistive signature for an AFM transition continues to pressures $p$$>$0.48 GPa [Fig.~2(a)]. This evolution is seen more clearly in $d\rho/dT$ that is plotted in Fig.~2(b). At ambient pressure, $d\rho/dT$ peaks near 2.5 K, which is much lower than the N\'{e}el temperature $T_N$$\approx$3.5 K. From magnetic neutron diffraction on a sample with Ir content near $x$=0.42, large-moment incommensurate antiferromagnetism develops at 3.5 K and coexisting small-moment, commensurate AFM appears below 2.7 K\cite{Llobet-CeRhIn5_IrNeu}. Specific heat measurements on our crystal in zero field [left inset to Fig.~2(b)] find an inflection point in $C/T$ at 3.5 K and a peak near 2.5 K, mimicking the features in $d\rho/dT$. We, therefore, identify the inflection temperature in $d\rho/dT$ with the onset of incommensurate AFM and the peak as an approximation to the appearance of commensurate order. Applying a 9-T magnetic field retains these anomalies in $C/T$, showing that two magnetic transitions persist with nearly identical ordering temperatures as at zero field [right inset in Fig.~2(b)]. It is obvious from inspection of the $d\rho/dT$ curves that they narrow quickly with increasing pressure and that for 0.61$\leq$$p$$<$1.06 GPa the peak position changes much more slowly. At 1.06 GPa, $d\rho/dT$ approaches a constant at low temperatures, signaling that $\rho(T)$ at this pressure is almost a linear function of temperature and that a well-defined signature for an AFM transition has disappeared. This is seen more clearly in the inset to Fig.~2(a) where $\rho(T)$ at 1.06 GPa is essentially $T$-linear below about 0.75 K and there is no detectable evidence for an AFM transition.

In the right inset to Fig.~2(b), we display the results of AC calorimetry ($C_{ac}$) measurements on the same crystal but from a separate pressure run. At 0.38 GPa, signatures of both AFM and SC transitions can be identified from the temperature dependence of $C_{ac}/T$ at zero field, and the latter is washed out in the presence of $B$=9 T, consistent with resistivity measurements. For $p$=0.79 GPa and zero field, only the SC transition is visible, but when a 9-T magnetic field is turned on and SC is completely suppressed, there is kink in $C_{ac}/T$ near 0.76 K, as in CeCu$_{(6-x)}$Au$_x$\cite{Lohneysen-CeCu6Au2001}, that reflects an AFM transition that also is evident in $d\rho/dT$ for 0.61$<$$p$$<$1.06 GPa. Indeed, this small anomaly in $C_{ac}/T$ disappears when pressure reaches 1.09 GPa. These specific heat and resistivity results suggest that this magnetic order, hidden by SC at zero field, is of bulk nature, and that a quantum criticality is approached near 1.06 GPa.

Further evidence for criticality comes from fitting the resistivity $\rho(T)$ to a power law:
\begin{equation}
\rho(T)=\rho_0+\Delta\rho(T)=\rho_0+AT^n,
\label{Eq.1}
\end{equation}
where $\rho_0$ is the residual resistivity. As also shown in the inset to Fig.~2(a), the low-temperature resistivity at 2.20 GPa follows a quadratic temperature dependence below about 1 K and is typical of a Fermi liquid. We take this pressure as a reference for normalizing lower pressure data in the false-color contour plot in Fig.~2(c). In the low temperature limit, this plot shows that the residual resistivity peaks near 0.6 GPa, and this is characteristic of  scattering amplified by quantum fluctuations, see also Fig.~2(e). In Fig.~2(c), colored solid triangles denote $T_N$, the temperature from Fig.~2(b) where large-moment incommensurate AFM develops. The coincidence of strongest scattering and an extrapolation of this N\'{e}el temperature to $T$=0 suggests that there is a quantum-phase transition of this order near $p_{c1}$ = 0.6 GPa. This conclusion is supported further by the pressure-induced collapse of the broad maximum in $d\rho/dT$ to a much narrower peak at 0.61 GPa.  Nevertheless, the narrow peak persists to 1.06 GPa, indicating that another magnetic transition remains to this higher pressure. As we will demonstrate, magnetic order in this pressure range is a SDW, but neutron diffraction or nuclear quadrupole resonance (NQR) experiments are needed to determine the detailed nature of this order. Irrespective of the precise nature of the magnetism, the important point is that it exists in this pressure range and we label this magnetic transition $T_{m}$. Figure~2(d) presents a contour plot of the local exponent $n$ as a function of $p$ and $T$. Here, $n$ is derived from the local derivative, $n$=$d(\ln\Delta\rho)/d(\ln T)$. Though the residual resistivity is a maximum near 0.6 GPa, there is only a limited temperature range above this pressure where the resistivity exhibits NFL behavior with $n$$\sim$1.0. This may be due to the presence of magnetism below $T_{m}$. In contrast, residual scattering is not so enhanced but there is an extended temperature range with $n$$\sim$1.0 around $p_{c2}$=1.06 GPa where $T_{c}^{on}$ in zero field is a maximum. Interestingly, there is a substantially increased inelastic scattering rate manifested by peaks in the $A$ coefficient at both $p_{c1}$ and $p_{c2}$ [see Fig.~2(e)]. The maxima of $A$ are about 4$\mu\Omega\cdot$cm/K$^n$, comparable to that in CeRhIn$_5$ at $P_2$\cite{Knebel-CeRhIn5R}, indicating that the effective-mass enhancements are similar in these two systems. This comparison is shown in Fig. S1 and discussed further in the \textbf{Supplementary Information} (\textbf{SI}).

The apparent dichotomy of signatures for quantum criticality at $p_{c1}$ and $p_{c2}$ is a first indication that these QCPs may be different in character, and this is further supported through thermopower measurements. The thermopower is given by\cite{Ziman-Solid}
\begin{equation}
S\equiv\frac{\alpha}{\sigma}=\frac{\pi^2}{3}\frac{k_B^2T}{q}\frac{\partial\ln\sigma(\varepsilon)}{\partial\varepsilon}\bigg|_{\varepsilon=\varepsilon_F},
\label{Eq.2}
\end{equation}
where $k_B$ is Boltzman's constant, $q$ is the charge of carriers, $\sigma$ is the electrical conductivity, $\alpha$ is the Peltier conductivity, and $\varepsilon_F$ is the chemical potential at $T$=0. Being the energy-derivative of $\ln\sigma(\varepsilon)$, $S$ is more sensitive to the Fermi-surface topology than $\sigma$. Moreover, because the entropy current $\mathbf{J}_{S}$=$\frac{1}{T}\boldsymbol{\alpha}\cdot\mathbf{E}$ ($\mathbf{E}$ is electric field), thermopower provides a measure of transport entropy per each conduction carrier.

Figure~3(a) displays the temperature dependence of in-plane thermopower at selected pressures without an external magnetic field. Below $T_c^{on}$, $S(T)$ drops to 0, demonstrating an entropy-less SC ground state. $S$ changes sign from negative to positive near 2.6 K at $p$=0. With applied pressure, the temperature where the sign change occurs moves to lower temperatures, and at $p$=0.35 GPa, it coincides with $T_c^{on}$. Application of a 9-T field allows tracking the sign changes to lower temperatures, as shown in Fig.~3(b). At $p$=0.61 GPa, $S(T)$ remains positive down to 0.3 K, the base temperature of our measurements, but may become negative at even lower temperature. Additional measurements to lower temperatures would be useful to determine the sign of $S$ especially around this pressure. Irrespective of some uncertainty of $S(T)$ below 0.3 K, a sign change of thermopower with pressure seems inescapable [Fig.~3(c)]. A smooth extrapolation of $S(T)$ suggests that $S(T)$ remains positive to $T$$\rightarrow$0 when $p$$\geq$0.73 GPa. The initial slope of $S(T)$ at our lowest temperatures is plotted in Fig.~2(f) as a function of $p$. There are two notable features: a discontinuous jump in $S/T$ near $p_{c1}$ and a maximum at $p_{c2}$. With $S$ being sensitive to Fermi-surface topology, the jump in $S/T$ implies a qualitative change in electronic properties at $p_{c1}$.

The temperature dependence of $S/T$ around $p_{c1}$  and $p_{c2}$ is displayed in Figs.~4(a) and 4(b). For $p$$<$$p_{c1}$, {\it e.g.} 0.48 GPa, $S/T$ initially increases with decreasing $T$ and forms a broad peak before dropping sharply as $T$$\rightarrow$0. The maximum and sharp drop in $S/T$ move to lower temperatures for $p$=0.61 GPa, but at a slightly higher pressure (0.73 GPa), $S/T$ increases monotonically with decreasing temperature and tends to saturate to a finite value as $T$ approaches zero. In contrast to the asymmetry around $p_{c1}$, $S/T$ approaches $T$=0 symmetrically about $p_{c2}$.
% and tends to diverge in the limit of zero temperature, though it may roll over to a constant at temperatures below 0.3 K
To the extent that $S/T$ also is proportional to the Sommerfeld coefficient of specific heat $\gamma$ in a multi-band system\cite{Behnia-Thermopower}, this temperature evolution of $S/T$ around $p_{c2}$ is a clear signature of quantum criticality, as discussed in the following section.\\

\textbf{DISCUSSION}

Before comparing these observations to theoretical predictions of quantum criticality, we consider possible alternative interpretations for the abrupt jump and sign change of $S/T$ around $p_{c1}$. These possibilities include: (\rmnum{1}) a change in crystal fields, (\rmnum{2}) magnetic breakdown, (\rmnum{3}) a valence transition and (\rmnum{4}) and a Lifshitz transition. The crystal-field ground state of the CeRh$_{1-x}$Ir$_x$In$_5$ series is a $\Gamma_7$ for all compositions\cite{Willers-CeMIn5PNAS} and the energy difference between ground and first-excited states only {\it decreases} slightly from 6.9 meV in CeRhIn$_5$ to 6.7 meV in CeIrIn$_5$\cite{Christianson-Ce115CEF}. It is highly improbable that a pressure of only 0.6 GPa could produce a sufficient change in crystal-field configuration  to induce a jump in $S/T$. Magnetic breakdown leads to partial reconstruction of the Fermi surface and, consequently, could provide a plausible scenario for the jump in $S/T$. High field dHvA studies of CeRhIn$_5$, however, are consistent with a lack of evidence for such an effect for fields below 30 T\cite{JiaoL-CeRhIn5B}, a field much higher than used in the present study. Though we cannot fully rule out the possibility of magnetic breakdown at $p_{c1}$, this scenario seems unlikely. Critical valence fluctuations have been proposed theoretically\cite{Watanabe-ValenceFluc} as one explanation for properties of CeRhIn$_5$ at its critical pressure of 2.35 GPa where there is an abrupt jump from small to large Fermi surfaces. There is, however, no  evidence so far as we know from magnetic susceptibility\cite{Pagliuso-CeRhIn5_Ir}, soft x-ray spectroscopy\cite{Willers-CeMIn5PNAS} or resonant X-ray-emission spectroscopy\cite{Yamaoka-Ce115Valence} for a valence change across the Ce(Rh,Ir)In$_5$ phase diagram. It again seems very unlikely that a small pressure of 0.61 GPa applied to our sample would induce critical valence fluctuations. Finally, we consider the possibility that a Lifshitz transition might account for transport and thermopower behaviors near $p_{c1}$. A Lifshitz transition, which does reconstruct the Fermi surface, under certain circumstances can produce a jump and sign change in $S/T$ as a function of some non-thermal control parameter that tunes the chemical potential\cite{Buhmann-Lifzhitz} or magnetic exchange\cite{Kuromoto-Lifshitz}. Though these theoretical models\cite{Buhmann-Lifzhitz,Kuromoto-Lifshitz} may capture aspects of our experimental observations, presently it is not possible to compare directly predictions of these models to our results as a function of pressure. In contrast to these plausible interpretations, evidence presented below allows a more straightforward and compelling interpretation of our observations within the framework of quantum criticality.

A model of Kondo-breakdown and SDW QCPs anticipates the behaviors we find around $p_{c1}$ and $p_{c2}$\cite{Kim-S2010}. This theory predicts a strong increase in $S/T$ as $T$ goes to zero following an $a$$-$$bT^{0.5}$ law and that this increase is symmetric about a SDW QCP as it is approached from AFM and paramagnetic states, just as we find at $p_{c2}$ [Fig.~4(a) and 4(g)]. We, therefore, identify the magnetic order below $T_m$ as being a spin-density wave. At a Kondo-breakdown QCP, however, $S/T$ should be asymmetric about the QCP [Fig.~4(c)], in agreement with experimental results at $p_{c1}$ [Fig.~4(a)]. In this theory\cite{Kim-S2010,Kim-S2011}, a sharp peak in $S/T$ on the AFM side of a Kondo-breakdown QCP is expected and signals FS reconstruction. Above the peak, $S/T$ is predicted to follow a $T^{-1/3}$ (2D) or $-\log(T/T_0)$ (3D) temperature dependence. In our case, $S/T$ is better fitted by the latter [cf. Fig.~4(e-f)]. Such a $T$ dependence of $S/T$ also is found in YbRh$_2$Si$_2$ at $B_c$=65 mT\cite{Hartmann-YRSS} where an abrupt change in thermopower is accompanied by a field-induced jump in Fermi surface, implied from Hall effect measurements, that signals a Kondo-breakdown QCP\cite{Paschen-YbRh2Si2Hall}. We should note that a modified SDW-criticality theory incorporating strong coupling\cite{Abrahams-PNAS2012}, which predicts $S/T$$\propto$$T^{-1/4}$, also fails to describe our results at $p_{c1}$ [Fig.~4(e)]. Though the maximum in $S/T$ at 0.61 GPa is not as sharp as predicted theoretically, some rounding of the theoretically sharp feature is expected because of the multi-sheeted Fermi surface\cite{Shishido-Ce115FS} and the presence of disorder scattering, neither of which is included in this idealized model. Nevertheless, agreement between experiment and theory at both $p_{c1}$ and $p_{c2}$ is appealing [see Fig.~4(a-d)] and evidence that the two critical points are likely different in nature, $p_{c1}$ being a Kondo-breakdown QCP and $p_{c2}$ a SDW QCP. These results provide an example where two qualitatively different QCPs appear to be realized in a single material driven by a single ``clean" tuning parameter that does not break symmetry or induce spin-polarization.

These observations lead us to consider a so-called ``global" model of quantum criticality that predicts a sequence of two zero-temperature phase boundaries as a function of some non-thermal tuning parameter\cite{SiQ-PhysB2006,Coleman-JLTP2010,Nica-QCP2016}, which in our case is pressure. Like our experiments in a 9-T field, this model does not consider explicitly the possibility of superconductivity that theoretically can develop from fluctuations around both Kondo-breakdown\cite{Pixley-PRB2015} and SDW\cite{Monthoux-Nature2007} critical points. As a function of the tuning parameter, there is in the model a boundary between a magnetically ordered state with small FS (AFM$_{\text{S}}$) and a SDW state with large Fermi surface (AFM$_{\textbf{L}}$). As the tuning parameter is increased further, this QCP is followed by another  $T$=0 boundary between SDW and paramagnetic states (PM$_{\text{L}}$), see Fig.~S2 in \textbf{SI}. This sequence of quantum-phase transitions seems to be realized in CeRh$_{0.58}$Ir$_{0.42}$In$_5$. Indeed, a more general view of our key observation is that the isothermal $S/T$, in the low temperature limit, undergoes a sudden negative-to-positive jump across $p_{c1}$ [Figs.~2(f) and 4(a)]. This is consistent with a sudden change of the FS from small to large across an unconventional QCP of the Kondo-breakdown type at $p_{c1}$. We note that $S$ at low temperature is negative in CeRhIn$_5$ (data not shown) but positive in CeIrIn$_5$\cite{LuoY-CeIrIn5Nernst}. This change in sign reflects their different electronic structures: in the former, there are three electron FS sheets ($\alpha_{1,2,3}$, band-15) and one hole FS sheet ($\beta_2$, band-14)\cite{Shishido-CeRhIn5dHvA}, but the latter compound has an additional hole FS sheet ($\beta_1$, band-14)\cite{HagaY-CeIrIn5dHvA} as well as larger overall FS.

From the magnetic phase boundaries in Fig.~2(d), we also conclude that the FS reconstruction is accompanied by a boundary between large-moment antiferromagnetism and a SDW, as it is in CeRhIn$_5$ at very high magnetic fields\cite{JiaoL-CeRhIn5B}. If only part of the FS reconstructs at $p_{c1}$, it is possible that the SDW manifested at pressures between $p_{c1}$ and $p_{c2}$ is the small-moment, commensurate AFM order at atmospheric pressure that coexists with SC\cite{Llobet-CeRhIn5_IrNeu}, but if the change in electronic structure is more severe, the SDW at high pressures could be different. Without the possibility of dHvA measurements on CeRh$_{0.58}$Ir$_{0.42}$In$_5$, determining the symmetry of the SDW at $p_{c1}$$\leq$$p$$\leq$$p_{c2}$ would provide insight on how the Fermi surface changes at $p_{c1}$. In this regard, it is worth remarking that the transitions to both large-moment incommensurate order and the small-moment commensurate order go to zero simultaneously as a function of $x$ in CeRh$_{1-x}$Ir$_x$In$_5$ at atmospheric pressure\cite{Llobet-CeRhIn5_IrNeu}, unlike the pressure response of CeRh$_{0.58}$Ir$_{0.42}$In$_5$.

These models that predict the variation of thermopower around QCPs and the global phase diagram of criticality in heavy-fermion systems are both based on the concept of Kondo breakdown. The theoretically predicted thermopower, however, is for a Kondo-breakdown transition without incorporating any magnetic order\cite{Kim-S2010,Kim-S2011}. This model allows but does not necessarily require criticality of magnetic order simultaneous with a localization/delocalization transition of the $f$-electron and associated jump in Fermi-surface volume. In contrast, the transition across $p_{c1}$ in pressurized CeRh$_{0.58}$Ir$_{0.42}$In$_5$ is between two magnetically ordered phases, as described by the AFM$_{\text{S}}$-to-AFM$_{\text{L}}$ transition in the global model of criticality (Fig.~S2, \textbf{SI}). This suggests that the critical electronic properties of the transition in the magnetic background are indeed dominated by the destruction of Kondo effect. It would be very instructive to carry out similar experiments at a direct transition between AFM$_{\text{S}}$ and PM$_{\text{L}}$ phases, where destruction of the Kondo effect is concurrent with the onset of AFM order; such a setting arises in CeRhIn$_5$ under pressure\cite{Park-CeRhIn5QCP,Shishido-CeRhIn5dHvA}. Thus, our work not only brings new understanding about unconventional quantum criticality but also opens an important means to shed new light on the global phase diagram.\\

\textbf{CONCLUSIONS}

In summary, pressure-dependent resistivity and thermopower measurements of heavy-fermion CeRh$_{0.58}$Ir$_{0.42}$In$_5$ are consistent with two QCPs accessed in a single material with a single clean tuning parameter. As discussed, a straightforward and likely interpretation of experiments is that there is a Kondo-breakdown critical point at $p_{c1}$=0.6 GPa where the Fermi surface suddenly reconstructs and is accompanied with a transition from large-moment antiferromagnetism to SDW order. This QCP is followed at higher pressures by a SDW QCP at $p_{c2}$=1.06 GPa beyond which there is a heavy Fermi-liquid state. Thermopower measurements around these critical points provide an experimental manifestation of theoretical predictions. Our work, therefore, also suggests that thermopower can be applicable to detect a Fermi-surface change in other systems when a direct Fermi-surface measurement is not possible. Further, the experimentally established sequence of QCPs and their natures are anticipated in the global phase diagram of heavy-fermion quantum criticality that predicts a series of AFM$_{\textbf{S}}$-AFM$_{\textbf{L}}$-PM$_{\text{L}}$ transitions at zero-temperature as found in CeRh$_{0.58}$Ir$_{0.42}$In$_5$ and depicted in Fig.~S2 in \textbf{SI}. It seems likely that the criticality in CeRh$_{0.58}$Ir$_{0.42}$In$_5$ generalizes to other members in this series and underlies their superconductivity. These studies uncover systematic insights that should be applicable generally to understanding quantum criticality in heavy-fermion materials and more broadly to bad metals with strong correlations.\\

\textbf{METHODS}

Single crystalline CeRh$_{0.58}$Ir$_{0.42}$In$_5$ was grown from an Indium-rich flux that contained the target ratio of Ce:Rh:Ir \cite{Pagliuso-CeRhIn5_Ir,Petrovic-CeIrIn5HFSC}. The Ir concentration was confirmed by comparing $^{115}$In NQR spectra to previous measurements\cite{ZhengG-CeRhIn5_IrNQR} and by energy dispersive x-ray spectroscopy (EDX), both of which gave $x$=0.42(3). Though EDX showed that the Ir concentration was highly uniform, we cannot rule out small variations in Rh:Ir ratio throughout the crystal's bulk. Thermopower measurements were carried out by means of a steady-state technique\cite{LuoY-CeIrIn5Nernst}. Both electrical and thermal currents were applied along the \textbf{a}-axis that is also the direction of the external magnetic field. Heat capacity under pressure was measured by an AC calorimetric
method. A hybrid piston-clamp type cell, with Daphne 7373 as the pressure medium, generated hydrostatic pressures to 2.20 GPa. Pressure in the cell was determined from the superconducting transition of Pb. \\

Data availability\\
The authors declare that all source data supporting the findings of this study are available within the paper. \\

\textbf{ACKNOWLEDGEMENTS}

We thank T. Park, F. Ronning, J. Lawrence,  H. v. L\"{o}hneysen, J. Singleton, C. P\'epin and L. Jiao for insightful conversations and J. L. Sarrao for providing the sample used in this study. Work at Los Alamos was performed under the auspices of the U.S. Department of Energy, Division of Materials Sciences and Engineering. A. P. Dioguardi and P. F. S. Rosa acknowledge Director's Postdoctoral Fellowships supported through the Los Alamos LDRD program. Work at Rice University was in part supported by the NSF Grant No. DMR-1611392, the ARO Grant No. W911NF-14-1-0525, and the Robert A. Welch Foundation Grant No. C-1411. X. Lu acknowledges the support from National Key R\&D Program of China (Grant No. 2017YFA0303101) and National Natural Science Foundation of China (Grant No. 11374257).\\

\textbf{AUTHOR CONTRIBUTIONS}

Y. L., E. D. B., Q. S. and J. D. T. conceived and designed the experiments. E. D. B., A. P. D. and P. F. S. R. characterized the crystals. Y. L. and X. L. performed the pressure measurements. Y. L., Q. S. and J. D. T. discussed the data, interpreted the results, and wrote the paper with input from all the authors.\\

\textbf{ADDITIONAL INFORMATION}\\
\textbf{Supplementary information} accompanies the paper on the {\it npj Quantum Materials} website (http://doi.org/10.1038/s41535-018-0080-9).\\
\textbf{Competing interests:} The authors declare that they have no competing financial interests.\\
\textbf{Publisher's note:} Springer Nature remains neutral with regard to jurisdictional claims in published maps and institutional affiliations.

\newpage
\textbf{Figures Legends:}

\begin{figure}[!h]
\hspace*{-10pt}
\includegraphics[width=17cm]{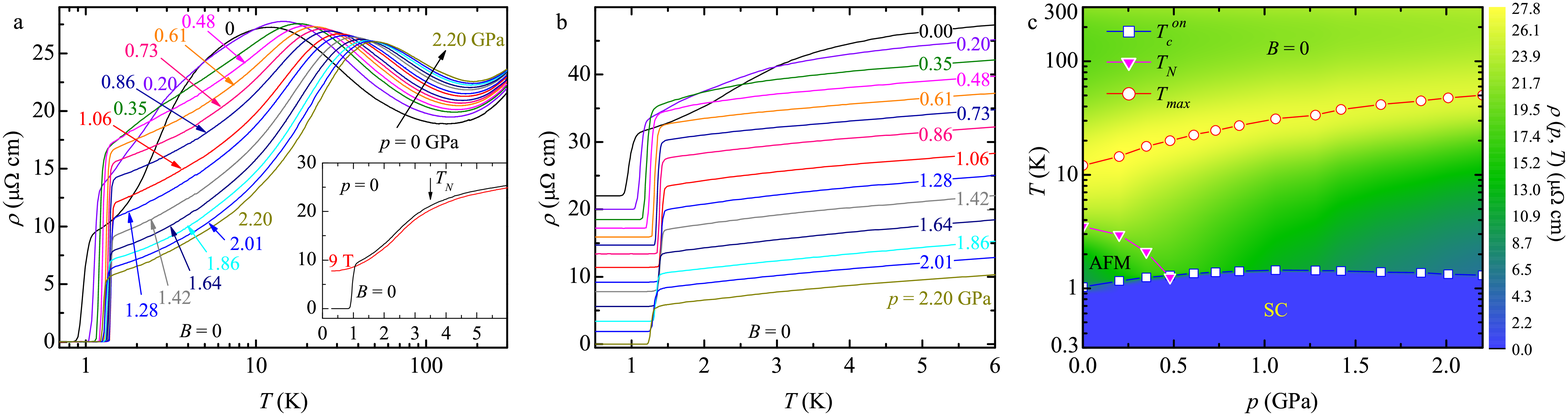}
\label{Fig1}
\end{figure}
\vspace*{-20pt}
\textbf{Fig.~1~} Resistivity of CeRh$_{0.58}$Ir$_{0.42}$In$_5$ under pressure. (a) Temperature-dependent resistivity of CeRh$_{0.58}$Ir$_{0.42}$In$_5$ under different pressures. The inset is a comparison of $\rho(T)$ for $B$=0 and 9 T at ambient pressure. (b) A zoom-in view of the low temperature region; the curves have been vertically shifted for clarity. (c) A contour plot constructed from $\rho(p, T)$ at zero magnetic field. Symbols show the evolution of resistivity maximum $T_{max}$, N\'{e}el temperature $T_{N}$ and superconducting onset temperature $T_{c}^{on}$. \\

\newpage

\begin{figure}[!h]
\hspace*{-15pt}
\vspace*{-13pt}
\includegraphics[width=16cm]{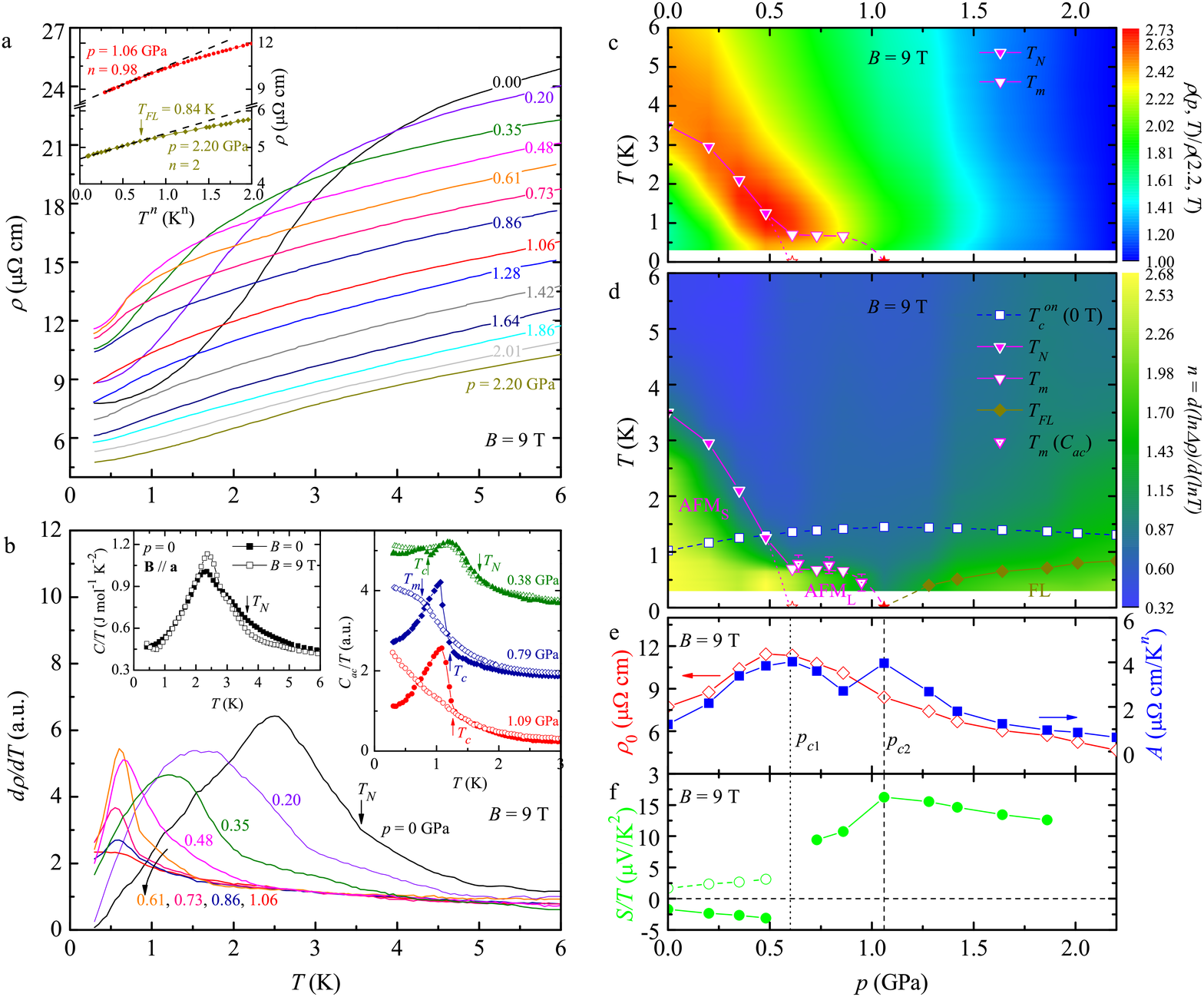}
\label{Fig2}
\end{figure}
\vspace*{-10pt}
\textbf{Fig.~2~} Transport properties of CeRh$_{0.58}$Ir$_{0.42}$In$_5$ at $B$=9 T, $\textbf{B}$$\parallel$$\textbf{a}$.
(a) $\rho(T)$ profiles at various pressures (data vertically shifted). The inset shows evidence for NFL and FL behaviors for $p$=1.06 GPa and 2.20 GPa, respectively. (b) $d\rho/dT$ for small applied pressures. The $\rho(T)$ data have been interpolated and three-point smoothed before differentiation. The left inset is the temperature-dependent specific heat divided by temperature ($C/T$) measured at $p$=0 for fields of 0 (solid) and 9 T (open). The upturn of $C/T$ below 0.7 K for 9 T is due to an In nuclear Schottky anomaly. The right inset plots AC calorimetry ($C_{ac}/T$) measured at 0.38, 0.79 and 1.09 GPa. The solid (open) symbols represent for $B$=0 (9 T). (c) Contour plot of isothermal resistivity normalized by the resistivity at 2.2 GPa, $\rho(p, T)/\rho(2.20, T)$. (d) False contour plot of the local exponent $n$=$d\ln(\Delta\rho)/d\ln(T)$ defined in Eq.~(\ref{Eq.1}). Symbols are defined in the legend and correspond to the properties obtained at 9 T, except for $T_c^{on}$. $T_m$ defined from AC calorimetry is also shown. (e) The residual resistivity $\rho_0$ (left axis) and coefficient $A$ (right axis). (f) The initial slope of thermopower, $S/T$, as a function of $p$; the open circles denote the absolute values for $p$$<$0.6 GPa.

\newpage

\begin{figure}[!h]
\hspace*{-10pt}
\includegraphics[width=17cm]{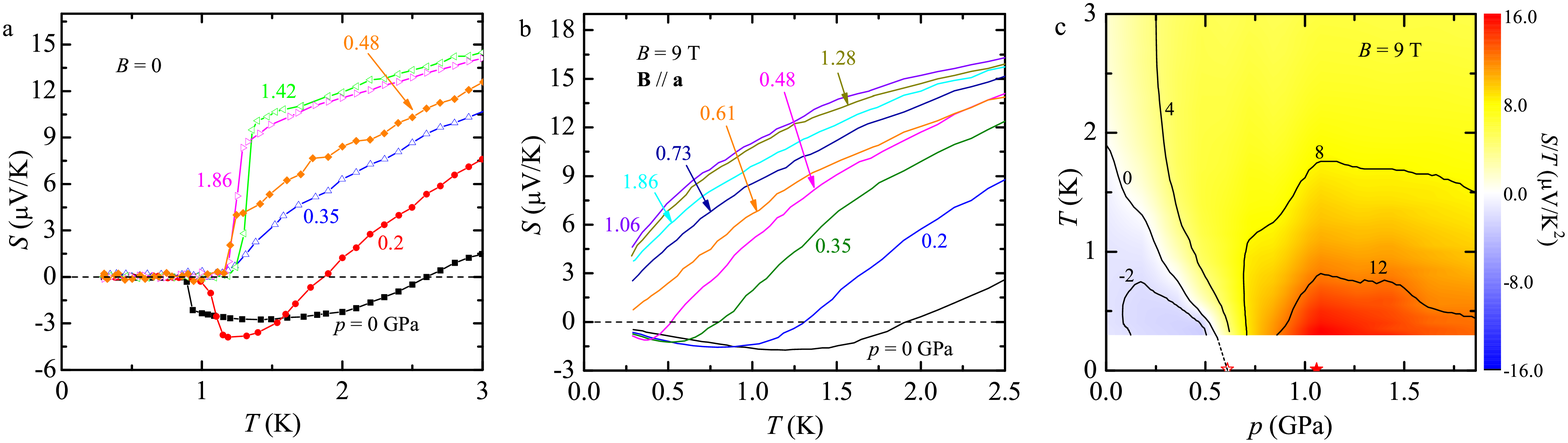}
\label{Fig3}
\end{figure}
\vspace*{-20pt}
\textbf{Fig.~3~} Thermopower of CeRh$_{0.58}$Ir$_{0.42}$In$_5$.
Temperature dependence of thermopower under various pressures, measured at zero magnetic field (a) and 9 T (b). (c) Contour plot of $S/T$ ($B$=9 T) as a function of $p$ and $T$. The boundary where $S/T$=0 extrapolates to near $p_{c1}$.  \\

\newpage

\begin{figure}[!h]
\hspace*{-5pt}
\includegraphics[width=16.5cm]{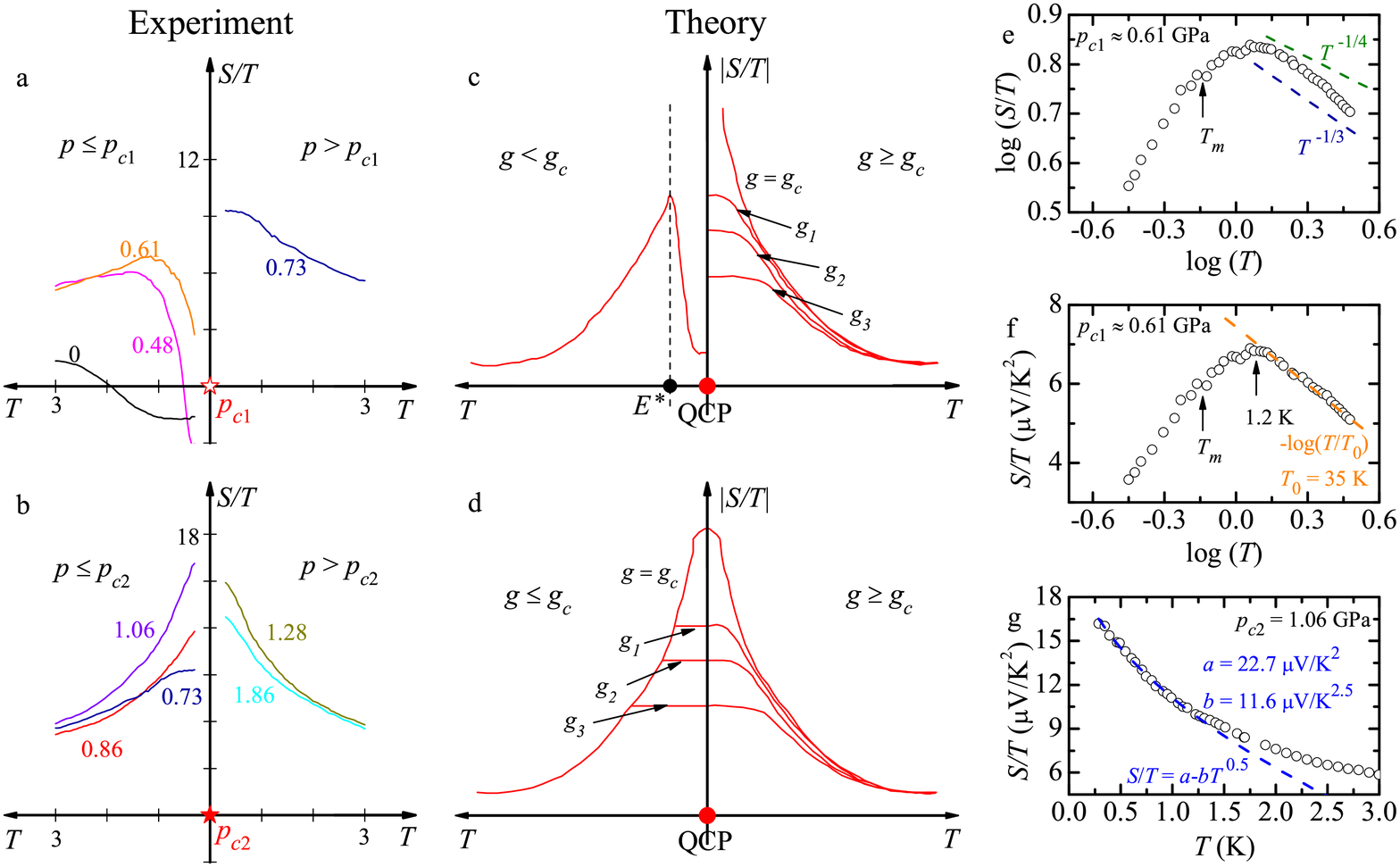}
\label{Fig3}
\end{figure}
\vspace*{-10pt}
\textbf{Fig.~4~} Thermopower near quantum critical points.
(a-d) A comparison of $S/T$ behaviors about a Kondo-breakdown QCP (upper panels) and a SDW QCP (lower panels). The left column shows experimental data (this work), and the right column shows the theoretical predictions reproduced from Ref.~\cite{Kim-S2010}. In the model calculations, a QCP appears at a critical value of a tuning parameter $g_c$. Values of $g_1$, $g_2$ and $g_3$ tune the system progressively away from a QCP. Note that theoretical curves are absolute values of $S/T$. (e) and (f) plot the temperature dependent $S/T$ near $p_{c1}$ and compare it to theoretically predicted dependencies for a Kondo-breakdown QCP as discussed in the text. The dashed line in (f) is a fit to the data above a maximum in $S/T$, with the fitting parameters indicated in the figure. (g) $S/T$ near $p_{c2}$ fitted to the theoretically predicted dependence $a$$-$$bT^{0.5}$ expected for a SDW QCP. \\

%%%%%%%%%%%%%%%%%%%%%%%%%%%%%%%%%%%%%%%%
%Starting of Supplementary Information
\newpage

\vspace{-15pt}

\begin{center}
{\it \textbf{Supplementary Information for: }} \\
\textbf{Unconventional and conventional quantum criticalities in CeRh$_{0.58}$Ir$_{0.42}$In$_5$}\\

\end{center}

\begin{center}
Yongkang Luo$^{1*}$\email{mpzslyk@gmail.com}, Xin Lu$^{2}$, Adam P. Dioguardi$^{1}$, Priscila F. S. Rosa$^{1}$, Eric D. Bauer$^{1}$, Qimiao Si$^{3}$ and Joe D. Thompson$^{1\dag}$\email{jdt@lanl.gov}\\
$^1${\it Los Alamos National Laboratory, Los Alamos, New Mexico 87545, USA;}\\
$^2${\it Center for Correlated Matter and Department of Physics, Zhejiang University, Hangzhou 310058, China; and}\\
$^3${\it Department of Physics and Astronomy and Center for Quantum Materials, Rice University, Houston, Texas 77005, USA.}

\end{center}

\vspace{10pt}

In this \textbf{Supplementary Information (SI)}, we provide additional comparison of $A$ coefficients for CeRhIn$_5$ and CeRh$_{0.58}$Ir$_{0.42}$In$_5$ that further supports the discussion and conclusions of the main text, and a perspective on the relationship between a global model of heavy-fermion quantum criticality and other members of the family to which CeRh$_{0.58}$Ir$_{0.42}$In$_5$ belongs. \\

\textbf{SI \Rmnum{1}. Comparison of $A$ coefficients for CeRhIn$_5$ and CeRh$_{0.58}$Ir$_{0.42}$In$_5$}

\begin{figure*}[!h]
\vspace*{-10pt}
\includegraphics[width=8.5cm]{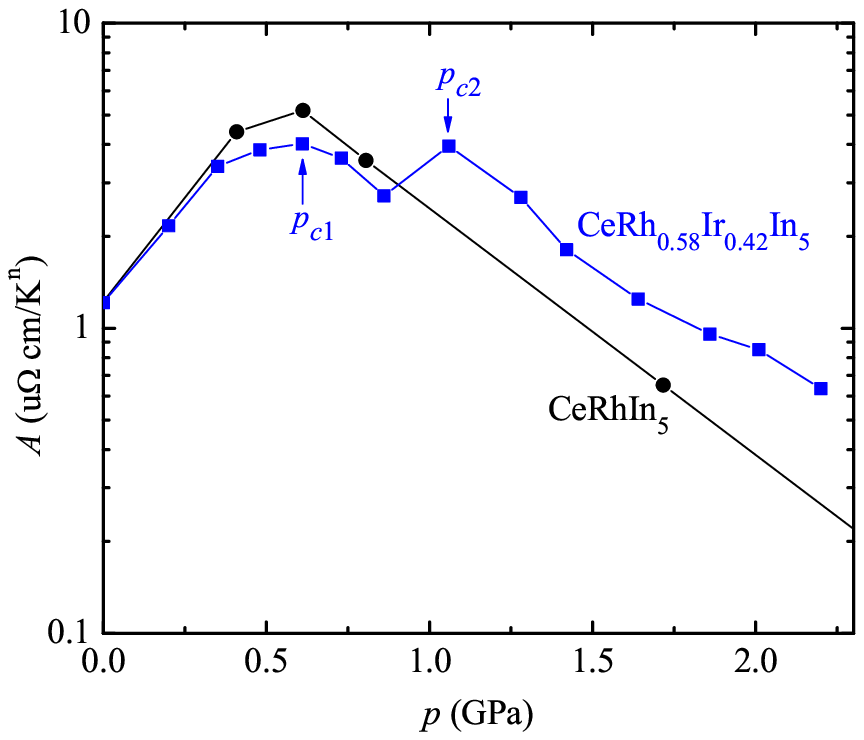}
\label{Fig.S1}
\end{figure*}
\vspace*{-10pt}
\textbf{Fig.~S1~} Comparison of $A$ coefficients of resistivity for CeRhIn$_5$ and CeRh$_{0.58}$Ir$_{0.42}$In$_5$ near their respective quantum critical points. Values of $A$ for CeRhIn$_5$ are taken from Ref.~\cite{SKnebel-CeRhIn5Pre}, and the pressure axis for CeRhIn$_5$ has been shifted properly for a more direct comparison to CeRh$_{0.58}$Ir$_{0.42}$In$_5$. \\

In a magnetic field sufficiently large to suppress pressure-induced superconductivity in CeRhIn$_5$, experiments find evidence for a magnetic quantum-critical point at $P_2$=2.4 GPa that is accompanied by a change from small-to-large Fermi surface\cite{SPark-CeRhIn5QCP,SShishido-CeRhIn5dHvA,SKnebel-CeRhIn5Pre}. Figure S1 plots the pressure dependence of the $A$ coefficient of resistivity, Eq.~(1) in the main text, of CeRhIn$_5$ as a function of pressure, where the pressure axis in this case has been shifted by about -1.8 GPa. For comparison, this figure includes the $A$ coefficient of CeRh$_{0.58}$Ir$_{0.42}$In$_5$ obtained in the present work. As shown in these plots, the evolution of $A(p)$ near the QCP for CeRhIn$_5$ is comparable to that of CeRh$_{0.58}$Ir$_{0.42}$In$_5$. In particular, the similarity in magnitude of $A$ coefficients at their QCPs, respectively ~5 and 4 $\Omega$$\cdot$cm/K$^n$, suggests comparably enhanced critical scattering and effective masses at their critical pressures. \\

\textbf{SI \Rmnum{2}. Quantum criticality and the larger family}

\begin{figure*}[!h]
\vspace*{5pt}
\includegraphics[width=8.5cm]{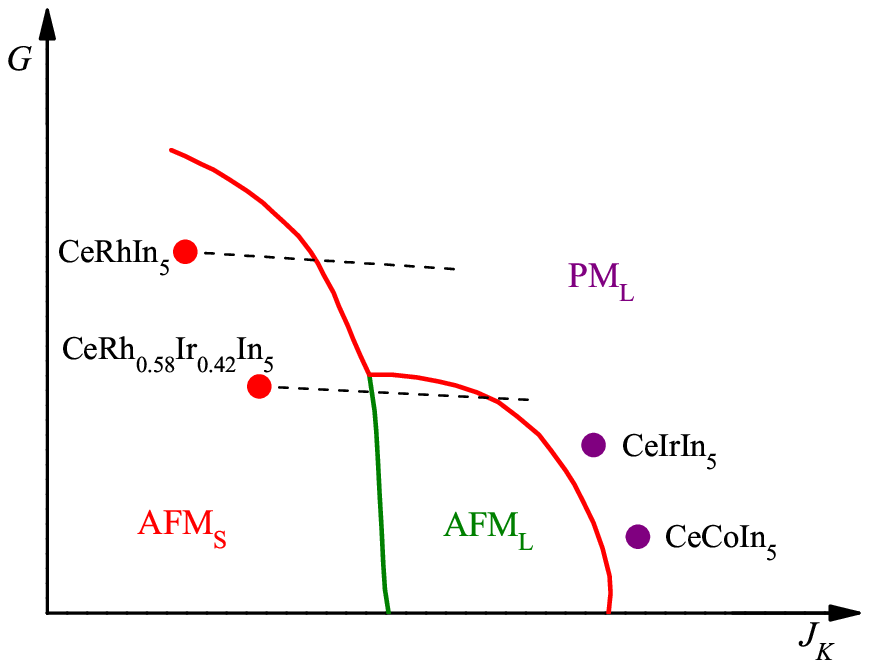}
\label{Fig.S2}
\end{figure*}
\vspace*{-10pt}
\textbf{Fig.~S2~} Ce$M$In$_5$ on the global magnetic phase diagram. $J_K$ is Kondo exchange, and $G$ labels frustration. The abbreviations are: AFM = antiferromagnetic and PM = paramagnetic. The subscript S (or L) denotes small (or large) Fermi surface.\\

A theoretically proposed global model of quantum criticality in heavy-fermion materials predicts a phase diagram illustrated schematically by the solid lines in Fig.~S2, where lines separate antiferromagnetic states with small (AFM$_{\text{S}}$) and large (AFM$_{\text{L}}$) Fermi surfaces and a paramagnetic state with large Fermi surface (PM$_{\text{L}}$). The lines have negative slope, reflecting the weakening of magnetism by both the magnetic frustration $G$ and Kondo exchange $J_K$\cite{SSiQ-PhysB2006,SColeman-JLTP2010}. Kondo-breakdown criticality occurs across the boundary between states with small and large Fermi surfaces\cite{SSiQ-PhysB2006}. Magnetic interactions in CeRhIn$_5$ are frustrated\cite{SDas-CeRhIn5}, the Kondo exchange is small and CeRhIn$_5$ crosses a Kondo-breakdown critical point as a function of pressure, leading to its placement in Fig.~S2. Hybridization and hence Kondo exchange increase in the sequence CeRhIn$_5$, CeIrIn$_5$ to CeCoIn$_5$\cite{SWillers-CeMIn5PNAS}, with CeIrIn$_5$ and CeCoIn$_5$ being very close to a AFM$_{\text{L}}$ QCP at atmospheric pressure\cite{SHarrison-CeMIn5dHvA,SThompson-Ce115JPSJ}. Consequently, they are placed on the diagram as shown. The sequence of Kondo-breakdown and then SDW critical points in CeRh$_{0.58}$Ir$_{0.42}$In$_5$ suggests its location in this diagram. Because pressure increases Kondo exchange, CeRh$_{0.58}$Ir$_{0.42}$In$_5$ as a function of pressure should have the trajectory indicated by the dashed line.  Our results on CeRh$_{0.58}$Ir$_{0.42}$In$_5$ suggest this broader perspective on quantum criticality in this larger family of materials.


\begin{thebibliography}{43}%
\bibitem{Coleman-QCP2005}
Coleman, P. \& Schofield, A. J. Quantum criticality. {\it Nature} \textbf{433}, 226-229 (2005).
\bibitem{Gegenwart2008}
Gegenwart, P., Si, Q. \& Steglich, F. Quantum criticality in heavy-fermion metals. {\it Nat. Phys.} \textbf{4}, 186-197 (2008).
\bibitem{Sachdev-QPT}
Sachdev, S. {\it Quantum Phase Transitions.} (Cambridge University Press, 2001).
\bibitem{Lohneysen-RMP2007}
L{\"o}hneysen, H.~v., Rosch, A., Vojta, M. \& W{\"o}lfle, P. Fermi-liquid instabilities at magnetic quantum phase transitions. {\it Rev. Mod. Phys.} \textbf{79}, 1015-1075 (2007).
\bibitem{Hertz-QCP}
Hertz, J. A. Quantum critical phenomena. {\it Phys. Rev. B} \textbf{14}, 1165-1184 (1976).
\bibitem{Millis-QCP}
Millis, A. J. Effect of a nonzero temperature on quantum critical points in itinerant fermion systems. {\it Phys. Rev. B} \textbf{48}, 7183-7196 (1993).
\bibitem{Schroder-CeCu6AuQCP}
Schr\"{o}der, A. {\it et al.} Onset of antiferromagnetism in heavy-fermion metals. {\it Nature} \textbf{407}, 351-355 (2000).
\bibitem{Custers-YbRh2Si2QCP}
Custers, J. {\it et al.} The break-up of heavy electrons at a quantum critical point. {\it Nature} \textbf{424}, 524-527 (2003).
\bibitem{Friedemann-YbRh2Si2_CoIr}
Friedemann, S. {\it et al.} Detaching the antiferromagnetic quantum critical point from the Fermi-surface reconstruction in YbRh$_2$Si$_2$. {\it Nat. Phys.} \textbf{5}, 465-469 (2009).
\bibitem{Custers-Ce3Pd20Si6QCP}
Custers, J. {\it et al.} Destruction of the Kondo effect in the cubic heavy-fermion compound Ce$_3$Pd$_{20}$Si$_6$. {\it Nat. Mater.} \textbf{11}, 189-194 (2012).
\bibitem{Paschen-YbRh2Si2Hall}
Paschen, S. {\it et al.} Hall-effect evolution across a heavy-fermion quantum critical point. {\it Nature} \textbf{432}, 881-885 (2004).
\bibitem{LuoY-CeNiAsOQCP}
Luo, Y. {\it et al.} Heavy-fermion quantum criticality and destruction of the Kondo effect in a nickel oxypnictide. {\it Nat. Mater.} \textbf{11}, 777-781 (2014).
\bibitem{LuoY-CeNi2As2Pre}
Luo, Y. {\it et al.} Pressure-tuned quantum criticality in the antiferromagnetic Kondo semi-metal CeNi$_{2-\delta}$As$_2$. {\it Proc. Natl. Acad. Sci. USA} \textbf{112}, 13520-13524 (2015).
\bibitem{SiQ-localQCP}
Si, Q., Rabello, S., Ingersent, K. \& Smith, J. L. Locally critical quantum phase transitions in strongly correlated metals. {\it Nature} \textbf{413}, 804-808 (2001).
\bibitem{Pepin-Mott2008}
P\'epin, C. Selective Mott transition and heavy fermions. {\it Phys. Rev. B} \textbf{77}, 245129 (2008).
\bibitem{Lohneysen-CeCu6Au2001}
L\"{o}hneysen, H. v., Pfleiderer, C., Pietrus, T., Stockert, O. \& Will, W. Pressure versus magnetic-field tuning of a magnetic quantum phase transition. {\it Phys. Rev. B} \textbf{63}, 134411 (2001).
\bibitem{Park-CeRhIn5QCP}
Park, T. {\it et al.} Hidden magnetism and quantum criticality in the heavy fermion superconductor CeRhIn$_5$. {\it Nature} \textbf{440}, 65-68 (2006).
\bibitem{Knebel-CeRhIn5Pre}
Knebel, G., Aoki, D., Braithwaite, D., Salce, B. \& Flouquet, J. Coexistence of antiferromagnetism and superconductivity in CeRhIn$_5$ under high pressure and magnetic field. {\it Phys. Rev. B} \textbf{74}, 020501 (2006).
\bibitem{Park-CeRhIn5NJP}
Park, T. \& Thompson, J. D. Magnetism and superconductivity in strongly correlated CeRhIn$_5$. {\it New J. Phys.} \textbf{11}, 055062 (2009).
\bibitem{Shishido-CeRhIn5dHvA}
Shishido, H., Settai, R., Harima, H. \& \={O}nuki, Y. A drastic change of the Fermi surface at a critical pressure in CeRhIn$_5$: dHvA study under pressure. {\it J. Phys. Soc. Jpn.} \textbf{74}, 1103-1106 (2005).
\bibitem{JiaoL-CeRhIn5B}
Jiao, L. {\it et al.} Fermi surface reconstruction and multiple quantum phase transitions in the antiferromagnet CeRhIn$_5$. {\it Proc. Natl. Acad. Sci. USA} \textbf{112}, 673-678 (2015).
\bibitem{Moll-CeRhIn5B}
Moll, P. J. W. {\it et al.} Field-induced density wave in the heavy-fermion compound CeRhIn$_5$. {\it Nat. Commun.} \textbf{6}, 6663 (2015).
\bibitem{Bao-CeRhIn5Neu}
Bao, W. {\it et al.} Incommensurate magnetic structure of CeRhIn$_5$. {\it Phys. Rev. B} \textbf{62}, R14621-R14624 (2000).
\bibitem{Llobet-CeRhIn5_IrNeu}
Llobet, A. {\it et al.} Novel coexistence of superconductivity with two distinct magnetic orders. {\it Phys. Rev. Lett.} \textbf{95}, 217002 (2005).
\bibitem{Nicklas-CeRhIn5_Ir}
Nicklas, M. {\it et al.} Two superconducting phases in CeRh$_{1-x}$Ir$_x$In$_5$. {\it Phys. Rev. B} \textbf{70}, 020505 (2004).
\bibitem{Pagliuso-CeRhIn5_Ir}
Pagliuso, P. G. {\it et al.} Coexistence of magnetism and superconductivity in CeRh$_{1-x}$Ir$_x$In$_5$. {\it Phys. Rev. B} \textbf{64}, 100503 (2001).
\bibitem{Aso-CeRhIn5Neu}
Aso, N. {\it et al.} Switching of magnetic ordering in CeRhIn$_5$ under hydrostatic pressure. {\it J. Phys. Soc. Jpn.} \textbf{78}, 073703 (2009).
\bibitem{Shishido-Ce115FS}
Shishido, H. {\it et al.} Fermi surface, magnetic and superconducting properties of LaRhIn$_5$ and CeTIn$_5$ (T: Co, Rh and Ir). {\it J. Phys. Soc. Jpn.} \textbf{71}, 162-173 (2002).
\bibitem{ZhengG-CeRhIn5_IrNQR}
Zheng, G. -q. {\it et al.} Coexistence of antiferromagnetic order and unconventional superconductivity in heavy-fermion CeRh$_{1-x}$Ir$_x$In$_5$ compounds: Nuclear quadrupole resonance studies. {\it Phys. Rev. B} \textbf{70}, 014511 (2004).
%\bibitem{Gotze-CeRh2Si2dHvA}
%G\"otze, K., Aoki, D., L\'evy-Bertrand, F., Harima, H. \& Sheikin, I. Drastic change of the Fermi surface across the metamagnetic transition in %CeRh$_2$Si$_2$. {\it Phys. Rev. B} \textbf{95} 161107 (2017).
\bibitem{Knebel-CeRhIn5R}
Knebel, G., Aoki, D., Brison, J. P., \& Flouquet, J. The Quantum Critical Point in CeRhIn$_5$: A Resistivity Study. {\it J. Phys. Soc. Jpn.} \textbf{77}, 114704 (2008).
\bibitem{Ziman-Solid}
Ziman, J. M. {\it Principles of the Theory of Solids.} (Cambridge University Press, 1972).
%\bibitem{Delves-Thermomagnetic}
%Delves, R. T. Thermomagnetic effects in semiconductors and semimetals. {\it Rep. Prog. Phys.} \textbf{28}, 249-289 (1965).
\bibitem{Behnia-Thermopower}
Behnia, K., Jaccard, D. \& Flouquet, J. On the thermoelectricity of correlated electrons in the zero-temperature limit. {\it J. Phys.: Condens. Matter} \textbf{16}, 5187-5198 (2004).
\bibitem{Willers-CeMIn5PNAS}
Willers, T. {\it et al.} Correlation between ground state and orbital anisotropy in heavy fermion materials. {\it Proc. Natl. Acad. Sci. USA} \textbf{112}, 2384-2388 (2015).
\bibitem{Christianson-Ce115CEF}
Christianson, A. D. {\it et al.} Crystalline electric field effects in Ce$M$In$_5$($M$=Co,Rh,Ir): Superconductivity and the influence of Kondo spin fluctuations. {\it Phys. Rev. B} \textbf{70}, 134505 (2004).
\bibitem{Watanabe-ValenceFluc}
Watanabe, S. \& Miyake, K. Origin of drastic change of Fermi surface and transport anomalies in CeRhIn$_5$ under pressure. {\it J. Phys. Soc. Jpn.} \textbf{79}, 033707 (2010).
\bibitem{Yamaoka-Ce115Valence}
Yamaoka, H. {\it et al.} Pressure and temperature dependence of the Ce valence and $c$-$f$ hybridization gap in Ce$T$In$_5$($T$=Co,Rh,Ir) heavy-fermion superconductors. {\it Phys. Rev. B} \textbf{92}, 235110 (2015).
%\bibitem{Lifshitz}
%Lifshitz, I. M. Soviet Physics JETP \textbf{11}, 1130 (1960).
\bibitem{Buhmann-Lifzhitz}
Buhmann, J. M. \& Sigrist, M. Thermoelectric effect of correlated metals: Band-structure effects and the breakdown of Mott's formula. {\it Phys. Rev. B} \textbf{88}, 115128 (2013).
\bibitem{Kuromoto-Lifshitz}
Kuromoto, Y. \& Hoshino, S. Composite Orders and Lifshitz Transition of Heavy Electrons. {\it J. Phys. Soc. Jpn.} \textbf{83}, 061007 (2014).
\bibitem{Kim-S2010}
Kim, K. S. \& P\'epin, C. Thermopower as a signature of quantum criticality in heavy fermions. {\it Phys. Rev. B} \textbf{81}, 205108 (2010).
\bibitem{Kim-S2011}
Kim, K. S. \& P\'epin, C. Thermopower as a fingerprint of the Kondo breakdown quantum critical point. {\it Phys. Rev. B} \textbf{83} 073104 (2011).
\bibitem{Hartmann-YRSS}
Hartmann, S. {\it et al.} Thermopower evidence for an abrupt Fermi surface change at the quantum critical point of YbRh$_2$Si$_2$. {\it Phys. Rev. Lett.} \textbf{104}, 096401 (2010).
\bibitem{Abrahams-PNAS2012}
Abrahams, E. \& W\"{o}lfle, P. Critical quasiparticle theory applied to heavy fermion metals near an antiferromagnetic quantum phase transition. {\it Proc. Natl. Acad. Sci. USA} \textbf{109} 3238-3242 (2012).
\bibitem{SiQ-PhysB2006}
Si, Q. Global magnetic phase diagram and local quantum criticality in heavy fermion metals. {\it Physica B} \textbf{378-380}, 23-27 (2006).
\bibitem{Coleman-JLTP2010}
Coleman, P. \& Nevidomskyy, A. H. Frustration and the Kondo effect in heavy fermion materials. {\it J. Low Temp. Phys.} \textbf{161}, 182-202 (2010).
\bibitem{Nica-QCP2016}
Nica, E. M., Ingersent, K. \& Si, Q. Quantum criticality and global phase diagram of an Ising-anisotropic Kondo lattice. {\it arXiv:} 1603.03829 (2016).
\bibitem{Pixley-PRB2015}
Pixley, J. H., Deng, L., Ingersent, K., \&  Si, Q. Pairing correlations near a Kondo-destruction quantum critical point. {\it Phys. Rev. B} \textbf{91}, 201109(R)(2015).
\bibitem{Monthoux-Nature2007}
Monthoux, P., Pines, D. \& Lonzarich, G. G. Superconductivity without phonons. {\it Nature} \textbf{450}, 1177 (2007).
\bibitem{LuoY-CeIrIn5Nernst}
Luo, Y., Rosa, P. F. S., Bauer, E. D. \& Thompson, J. D. Vortexlike excitations in the heavy-fermion superconductor CeIrIn$_5$. {\it Phys. Rev. B} \textbf{93}, 201102 (2016).
\bibitem{HagaY-CeIrIn5dHvA}
Haga, Y. {\it et al.} Quasi-two-dimensional Fermi surfaces of the heavy fermion superconductor CeIrIn$_5$. {\it Phys. Rev. B} \textbf{63}, 060503 (2001).
\bibitem{Petrovic-CeIrIn5HFSC}
Petrovic, C. {\it et al.} A new heavy-fermion superconductor CeIrIn$_5$: A relative of the cuprates? {\it EPL} \textbf{53} 354¨C359 (2001).


\end{thebibliography}

\begin{thebibliography}{43}%
%\bibitem{SKim-S2010}
%Kim, K. S. \& P\'epin, C. Thermopower as a signature of quantum criticality in heavy fermions. {\it Phys. Rev. B} \textbf{81}, 205108 (2010).
%\bibitem{SKim-S2011}
%Kim, K. S. \& P\'epin, C. Thermopower as a fingerprint of the Kondo breakdown quantum critical point. {\it Phys. Rev. B} \textbf{83} 073104 (2011).
\bibitem{SPark-CeRhIn5QCP}
Park, T. {\it et al.} Hidden magnetism and quantum criticality in the heavy fermion superconductor CeRhIn$_5$. {\it Nature} \textbf{440}, 65-68 (2006).
\bibitem{SShishido-CeRhIn5dHvA}
Shishido, H., Settai, R., Harima, H. \& \={O}nuki, Y. A drastic change of the Fermi surface at a critical pressure in CeRhIn$_5$: dHvA study under pressure. {\it J. Phys. Soc. Jpn.} \textbf{74}, 1103-1106 (2005).
\bibitem{SKnebel-CeRhIn5Pre}
Knebel, G., Aoki, D., Brison, J-P \& Flouquet, J. The quantum critical point in CeRhIn$_5$: a resistivity study. {\it J. Phys. Soc. Jpn.} \textbf{77}, 114704 (2008).
\bibitem{SSiQ-PhysB2006}
Si, Q. Global magnetic phase diagram and local quantum criticality in heavy fermion metals. {\it Physica B} \textbf{378-380}, 23-27 (2006).
\bibitem{SColeman-JLTP2010}
Coleman, P. \& Nevidomskyy, A. H. Frustration and the Kondo effect in heavy fermion materials. {\it J. Low Temp. Phys.} \textbf{161}, 182-202 (2010).
\bibitem{SDas-CeRhIn5}
Das, P. {\it et al.} Magnitude of the magnetic exchange interaction in the heavy-fermion antiferromagnet CeRhIn$_5$. {\it Phys. Rev. Lett.} \textbf{113}, 246403 (2014).
\bibitem{SWillers-CeMIn5PNAS}
Willers, T. {\it et al.} Correlation between ground state and orbital anisotropy in heavy fermion materials. {\it Proc. Natl. Acad. Sci. USA} \textbf{112}, 2384-2388 (2015).
\bibitem{SHarrison-CeMIn5dHvA}
Harrison, N. {\it et al.} $4f$-electron localization in Ce$_x$La$_{1-x}$$M$In$_5$ with $M$=Co, Rh, or Ir. {\it Phys. Rev. Lett.} \textbf{93}, 186405 (2004).
\bibitem{SThompson-Ce115JPSJ}
Thompson, J. D. \& Fisk, Z. Progress in heavy-fermion superconductivity: Ce115 and related materials. {\it J. Phys. Soc. Jpn.} \textbf{81}, 011002 (2012).
\end{thebibliography}
\end{document}